\documentclass[prd,twocolumn,twoside,preprintnumbers,superscriptaddress,nofootinbib]{revtex4}

\usepackage{amsmath,slashed}
\usepackage{graphicx,graphics,color}
\usepackage{dcolumn}
\usepackage[hyperfootnotes=false]{hyperref}
\usepackage{xspace}
\usepackage{enumerate}
\usepackage{varwidth}

\usepackage{physics}
\usepackage{amssymb}
\usepackage{mathtools}
\usepackage{dsfont}
\usepackage{multirow}

\newcommand{\mpii}{M_{\pi^0}}

\newcommand{\beq}{\begin{equation}}
\newcommand{\eeq}{\end{equation}}
\newcommand{\diff}{\text{d}}
\newcommand{\eps}{\epsilon}

\newcommand{\Order}{\mathcal{O}}

\newcommand{\mw}{M_\omega}
\newcommand{\Gw}{\Gamma_\omega}
\newcommand{\mr}{M_\rho}
\newcommand{\Gr}{\Gamma_\rho}

\newcommand{\epsrw}{\eps_{\omega}}
\newcommand{\gwg}{g_{\omega\gamma}}
\newcommand{\grg}{g_{\rho\gamma}}

\newcommand{\GeV}{\,\text{GeV}}
\newcommand{\MeV}{\,\text{MeV}}

\allowdisplaybreaks[1]

\begin{document}

\preprint{PSI-PR-23-22,  ZU-TH 32/23, IPARCOS-UCM-23-077}

\title{A phenomenological estimate of isospin breaking in hadronic vacuum polarization}

\author{Martin Hoferichter}
\affiliation{Albert Einstein Center for Fundamental Physics, Institute for Theoretical Physics, University of Bern, Sidlerstrasse 5, 3012 Bern, Switzerland}
\author{Gilberto Colangelo}
\affiliation{Albert Einstein Center for Fundamental Physics, Institute for Theoretical Physics, University of Bern, Sidlerstrasse 5, 3012 Bern, Switzerland}
\author{Bai-Long Hoid}
\affiliation{Albert Einstein Center for Fundamental Physics, Institute for Theoretical Physics, University of Bern, Sidlerstrasse 5, 3012 Bern, Switzerland}
\author{Bastian Kubis}
\affiliation{Helmholtz-Institut f\"ur Strahlen- und Kernphysik (Theorie) and
Bethe Center for Theoretical Physics, Universit\"at Bonn, 53115 Bonn, Germany}
\author{Jacobo Ruiz de Elvira}
\affiliation{Departamento de F\'isica Te\'orica and IPARCOS, Facultad de Ciencias F\'isicas, Universidad Complutense de Madrid, Plaza de las Ciencias 1, 28040 Madrid, Spain}
\author{Dominic Schuh}
\affiliation{Helmholtz-Institut f\"ur Strahlen- und Kernphysik (Theorie) and
Bethe Center for Theoretical Physics, Universit\"at Bonn, 53115 Bonn, Germany}
\author{Dominik Stamen}
\affiliation{Helmholtz-Institut f\"ur Strahlen- und Kernphysik (Theorie) and
Bethe Center for Theoretical Physics, Universit\"at Bonn, 53115 Bonn, Germany}
\author{Peter Stoffer}
\affiliation{Physik-Institut, Universit\"at Z\"urich, Winterthurerstrasse 190, 8057 Z\"urich, Switzerland}
\affiliation{Paul Scherrer Institut, 5232 Villigen PSI, Switzerland}

\begin{abstract} 
 Puzzles in the determination of the hadronic-vacuum-polarization contribution currently impede a conclusive interpretation of the precision measurement of the anomalous magnetic moment of the muon at the Fermilab experiment. One such puzzle concerns tensions between evaluations in lattice QCD and using $e^+e^-\to\text{hadrons}$ cross-section data. In lattice QCD, the dominant isospin-symmetric part and isospin-breaking (IB) corrections are calculated separately, with very different systematic effects. Identifying these two pieces in a data-driven approach provides an opportunity to compare them individually and trace back the source of the discrepancy. 
 Here, we estimate the IB component of the lattice-QCD calculations from phenomenology, based on a comprehensive study of exclusive contributions that can be enhanced via infrared singularities, threshold effects, or hadronic resonances, including, for the first time, in the $e^+e^-\to3\pi$ channel. We observe sizable cancellations among different channels, with a sum that even suggests a slightly larger result for the QED correction than obtained in lattice QCD. We conclude that the tensions between lattice QCD and $e^+e^-$ data therefore cannot be explained by the IB contributions in the lattice-QCD calculations.     
\end{abstract}

\maketitle

\emph{Introduction}.---The anomalous magnetic moment of the muon $a_\mu=(g-2)_\mu/2$ has been measured at a precision of 0.35 parts per million ~\cite{Muong-2:2021ojo,Muong-2:2021ovs,Muong-2:2021xzz,Muong-2:2021vma,Muong-2:2006rrc},
\begin{equation}
 a_\mu^\text{exp}=116\,592\,061(41)\times 10^{-11},
\end{equation}
and further improvements are expected shortly from the Fermilab experiment. To fully profit from this impressive achievement, a commensurate level of precision is required for the prediction within the Standard Model (SM)~\cite{Aoyama:2020ynm,Aoyama:2012wk,Aoyama:2019ryr,Czarnecki:2002nt,Gnendiger:2013pva,Davier:2017zfy,Keshavarzi:2018mgv,Colangelo:2018mtw,Hoferichter:2019gzf,Davier:2019can,Keshavarzi:2019abf,Hoid:2020xjs,Kurz:2014wya,Melnikov:2003xd,Colangelo:2014dfa,Colangelo:2014pva,Colangelo:2015ama,Masjuan:2017tvw,Colangelo:2017qdm,Colangelo:2017fiz,Hoferichter:2018dmo,Hoferichter:2018kwz,Gerardin:2019vio,Bijnens:2019ghy,Colangelo:2019lpu,Colangelo:2019uex,Blum:2019ugy,Colangelo:2014qya},
\begin{equation}
\label{amuSM}
a_\mu^\text{SM}[e^+e^-]=116\,591\,810(43)\times 10^{-11},
\end{equation}
whose uncertainty is dominated by hadronic effects. In particular, Eq.~\eqref{amuSM} reflects the prediction if the leading hadronic correction, hadronic vacuum polarization (HVP), is determined from $e^+e^-\to\text{hadrons}$ cross-section data, via the master formula~\cite{Bouchiat:1961lbg,Brodsky:1967sr} 
\begin{align}
\label{amu_HVP}
 a_\mu^\text{HVP}&=\bigg(\frac{\alpha m_\mu}{3\pi}\bigg)^2\int_{s_\text{thr}}^\infty \diff s \frac{\hat K(s)}{s^2}R_\text{had}(s),\notag\\
 R_\text{had}(s)&=\frac{3s}{4\pi\alpha^2}\sigma(e^+e^-\to\text{hadrons}(+\gamma)). 
\end{align} 
With the kernel function $\hat K(s)$ known in analytical form, the challenge amounts to a sub-percent-level measurement of the hadronic $R$-ratio $R_\text{had}$, including radiative corrections. By convention, the leading-order HVP contribution is defined photon-inclusively, in such a way that the integration threshold is determined by the $\pi^0\gamma$ channel, $s_\text{thr}=\mpii^2$. 

While for the subleading hadronic correction, hadronic light-by-light scattering, both subsequent lattice-QCD studies~\cite{Chao:2021tvp,Chao:2022xzg,Blum:2023vlm,Alexandrou:2022qyf,Gerardin:2023naa} and data-driven methods~\cite{Hoferichter:2020lap,Ludtke:2020moa,Bijnens:2020xnl,Bijnens:2021jqo,Zanke:2021wiq,Danilkin:2021icn,Colangelo:2021nkr,Holz:2022hwz,Leutgeb:2022lqw,Bijnens:2022itw,Ludtke:2023hvz,Hoferichter:2023tgp} have largely corroborated the evaluation from Ref.~\cite{Aoyama:2020ynm},\footnote{Further improvements required for the final Fermilab precision~\cite{Muong-2:2015xgu} are ongoing~\cite{Colangelo:2022jxc}, and also higher-order hadronic effects~\cite{Calmet:1976kd,Kurz:2014wya,Colangelo:2014qya,Hoferichter:2021wyj} are well under control.}
the HVP result underlying~\eqref{amuSM} has been challenged by lattice QCD
and the new $e^+e^-\to2\pi$ data by CMD-3~\cite{CMD-3:2023alj}.
For what concerns the latter, the question whether radiative corrections in the interpretation of $e^+e^-\to 2\pi$ data could provide an explanation of the discrepancy with other experiments is currently being investigated~\cite{Campanario:2019mjh,Colangelo:2022lzg,Ignatov:2022iou,JMPhDThesis,Abbiendi:2022liz}. 

Here, we instead focus on the comparison to lattice QCD. First, Ref.~\cite{Borsanyi:2020mff} differs for the entire HVP integral by $2.1\sigma$ from $e^+e^-$ data.
In addition, to try and disentangle different regions in center-of-mass energy $\sqrt{s}$, one can study 
so-called Euclidean-time windows~\cite{RBC:2018dos}, which amount to adding weight functions into the integral appearing in Eq.~\eqref{amu_HVP}.  With the parameter choices from Ref.~\cite{RBC:2018dos}, one typically defines a short-distance (SD), intermediate (int), and long-distance (LD) window, where the intermediate window behaves particularly well as regards systematic effects in lattice QCD. Accordingly, several collaborations have confirmed a significant tension in this quantity~\cite{Ce:2022kxy,ExtendedTwistedMass:2022jpw,FermilabLatticeHPQCD:2023jof,RBC:2023pvn,Colangelo:2022vok}, well beyond the level found in Ref.~\cite{Borsanyi:2020mff} for the full HVP contribution. At present, no simple solution to this puzzle is known, in particular in view of the pattern of changes in the intermediate window, total HVP, and the hadronic running of the fine-structure constant~\cite{Passera:2008jk,Crivellin:2020zul,Keshavarzi:2020bfy,Malaescu:2020zuc,Colangelo:2020lcg,Ce:2022eix},
and beyond-the-SM solutions become very contorted if at all possible~\cite{DiLuzio:2021uty,Darme:2021huc,Crivellin:2022gfu,Coyle:2023nmi}. 

\nocite{Borsanyi:2017zdw,Jegerlehner:2017gek,Hoferichter:2022iqe,Colangelo:2022prz,Colangelo:2023rqr,Hoferichter:2023bjm,Stamen:2022uqh}

\begin{table*}[t]
	\renewcommand{\arraystretch}{1.3}
	\centering
\begin{tabular}{crrrrrrrr}
\toprule
& \multicolumn{2}{c}{SD} & \multicolumn{2}{c}{int} & \multicolumn{2}{c}{LD} & \multicolumn{2}{c}{full}\\
& $\Order(e^2)$ & $\Order(\delta)$& $\Order(e^2)$ & $\Order(\delta)$& $\Order(e^2)$ & $\Order(\delta)$& $\Order(e^2)$ & $\Order(\delta)$\\\colrule
$\pi^0\gamma$ & $0.16(0)$ & -- & $1.52(2)$  & -- & $2.70(4)$ & -- & $4.38(6)$ & --\\
$\eta\gamma$ & $0.05(0)$ & -- & $0.34(1)$ & -- & $0.31(1)$ &-- & $0.70(2)$ & --\\
$\omega(\to\pi^0\gamma)\pi^0$ & $0.15(0)$ & -- & $0.54(1)$ & -- & $0.19(0)$ &-- & $0.88(2)$ & --\\\colrule
FSR ($2\pi$) & $0.12(0)$ & -- & $1.17(1)$ & -- & $3.13(3)$ &-- & $4.42(4)$ & --\\
FSR ($3\pi$) & $0.03(0)$ & -- & $0.20(0)$ & -- & $0.28(1)$ &-- & $0.51(1)$ & --\\
FSR ($K^+ K^-$)  & $0.07(0)$ & -- & $0.39(2)$ & -- & $0.29(2)$ &-- & $0.75(4)$ & --\\\colrule
$\rho$--$\omega$ mixing ($2\pi$) & -- & $0.06(1)$ & -- & $0.86(6)$ & -- & $2.87(12)$ & -- & $3.79(19)$\\
$\rho$--$\omega$ mixing ($3\pi$) & -- & $-0.13(3)$ & -- & $-1.03(27)$ & -- & $-1.52(40)$ & -- & $-2.68(70)$\\\colrule
pion mass ($2\pi$) & $0.04(8)$ & -- & $-0.09(56)$ & -- & $-7.62(63)$ & --& $-7.67(94)$ &--\\
kaon mass ($K^+ K^-$)  & $-0.29(1)$ & $0.44(2)$ & $-1.71(9)$ & $2.63(14)$ & $-1.24(6)$ & $1.91(10)$ & $-3.24(17)$ & $4.98(26)$\\
kaon mass ($\bar K^0 K^0$)  & $0.00(0)$ & $-0.41(2)$  & $-0.01(0)$ & $-2.44(12)$ & $-0.01(0)$ & $-1.78(9)$ & $-0.02(0)$ & $-4.62(23)$\\\colrule
sum  & $0.33(8)$ & $-0.04(4)$ & $2.34(57)$ & $0.02(33)$ & $-1.97(63)$ & $1.48(44)$ & $0.71(95)$ & $1.47(80)$\\
\botrule
\end{tabular}
\caption{IB effects to $a_\mu^\text{HVP}$ from the various radiative channels, FSR, $\rho$--$\omega$ mixing, and threshold effects, separated into $\Order(e^2)$ and $\Order(\delta)$ contributions as well as SD, intermediate, and LD windows. 
The $2\pi$ numbers for FSR and $\rho$--$\omega$ mixing from Ref.~\cite{Colangelo:2022prz} have been slightly updated to be consistent with the $\omega$ width extracted from the $3\pi$ channel~\cite{Hoferichter:2023bjm}. 
The uncertainties in the last line refer to the quadratic sum of the individual errors, see Table~\ref{tab:summary} for the additional uncertainty estimates discussed in the main text. All entries in units of $10^{-10}$.}  
	\label{tab:channels}
\end{table*}

Besides the consideration of windows, another point that can be scrutinized in lattice-QCD calculations concerns the separation into an isosymmetric part and isospin-breaking (IB) corrections, the latter further separated into QED, $\Order(e^2)$, and strong IB, $\Order(\delta=m_u-m_d)$, effects. Attempts to estimate such corrections from phenomenology have been made before~\cite{Borsanyi:2017zdw,Jegerlehner:2017gek,Hoferichter:2022iqe}, but with detailed work on the main exclusive channels, $e^+e^-\to 2\pi$~\cite{Colangelo:2018mtw,Colangelo:2022prz,Colangelo:2023rqr}, $3\pi$~\cite{Hoferichter:2019gzf,Hoferichter:2023bjm}, $\bar K K$~\cite{Stamen:2022uqh}, we are now in a position to evaluate all IB contributions for which enhancement from infrared (IR) singularities, threshold effects, or hadronic resonances is expected, and thus to provide a valuable point of comparison to lattice QCD for the IB corrections. This is the main purpose of this Letter.

\emph{Radiative channels}.---The first class of contributions arises from radiative channels, since the cross section entering the HVP integral~\eqref{amu_HVP} includes photonic corrections. By definition, it gives rise to a pure $\Order(e^2)$ effect, and the different channels can simply be evaluated in terms of their respective cross sections. For $e^+e^-\to\pi^0\gamma$ we use the dispersive analysis from Ref.~\cite{Hoid:2020xjs}, while for $\eta\gamma$ and $\omega(\to \pi^0\gamma)\pi^0$ a simple vector-meson-dominance model proves sufficient to reproduce the full evaluation from Refs.~\cite{Davier:2019can,Keshavarzi:2019abf}. For our analysis, we will use the values given in the first panel of Table~\ref{tab:channels}, where we also provide the decomposition onto Euclidean-time windows. Regarding the energy dependence of these contributions, $\pi^0\gamma$ and $\eta\gamma$ are dominated by $\omega$ and $\phi$ resonances, respectively, while the $\omega(\to\pi^0\gamma)\pi^0$ channel has a more complicated origin. In principle, it could be booked as a $\pi^0\pi^0\gamma$ final state, but since its contribution is qualitatively different from a radiative correction to $2\pi$, we follow the convention to list this channel separately. It arises via the transition form factor $\omega\to\pi^0\gamma^*$, coupling the virtual photon to the initial-state $e^+e^-$ pair, with a subsequent decay $\omega\to\pi^0\gamma$. Phenomenologically, this transition form factor is dominated by the $\rho(1450)$ resonance~\cite{Achasov:2013btb}, and thus leads to a surprisingly large radiative contribution  despite having its main support well beyond $1\GeV$. Other radiative channels, such as $\eta'\gamma$, are already negligible.

\begin{figure}[t]
 \includegraphics[width=\linewidth]{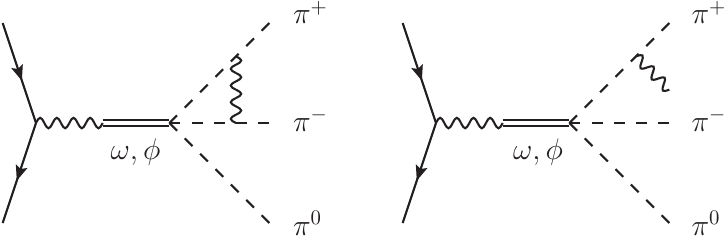}
 \caption{Representative diagrams for radiative corrections in $e^+e^-\to 3\pi$. IR-enhanced effects occur after the IR divergences from virtual photons (left) and real emission (right) have canceled. }
 \label{fig:3pidiagrams}
\end{figure}

\emph{Final-state radiation}.---The second class of contributions will be labeled as final-state radiation (FSR), even though it quantifies the combination of virtual-photon and bremsstrahlung diagrams. For the $2\pi$ channel, the effect is typically defined by an inclusive correction factor $\eta_{2\pi}$ ~\cite{Hoefer:2001mx,Czyz:2004rj,Gluza:2002ui,Bystritskiy:2005ib}, which sums up those contributions that inherit some IR enhancement after the IR singularities have canceled (see Refs.~\cite{JMPhDThesis,Radinprep} for a more general treatment of radiative corrections to $e^+e^-\to\pi^+\pi^-$). Indeed, it was shown in Ref.~\cite{Moussallam:2013una} that  the non-IR-enhanced $\pi^+\pi^-\gamma$ contribution (as well as $\pi^0\pi^0\gamma$ below $1\GeV$) is small, and a similar conclusion pertains to radiative corrections to the forward--backward asymmetry in $e^+e^-\to\pi^+\pi^-$~\cite{Ignatov:2022iou,Colangelo:2022lzg}. For the $2\pi$ channel, we thus use the outcome of Ref.~\cite{Colangelo:2022prz} based on $\eta_{2\pi}$, supplemented by the small corrections from Ref.~\cite{Moussallam:2013una}, and likewise for $K^+K^-$~\cite{Stamen:2022uqh}. For the $3\pi$ channel, 
the study of radiative corrections is far less advanced for two main reasons. First, the low-energy limit of the amplitude is related to the Wess--Zumino--Witten anomaly~\cite{Wess:1971yu,Witten:1983tw}, which leads to a non-renormalizable theory for radiative corrections~\cite{Ametller:2001yk,Ahmedov:2002tg,Bakmaev:2005sg}, while in the $2\pi$ channel the leading order in chiral perturbation theory (ChPT) coincides with scalar QED. Second, the non-radiative cross section is dominated by $\omega$ and $\phi$ resonances, which needs to be taken into account for any realistic estimate. Here, we focus again on the IR-enhanced radiative corrections, which arise from the diagrams shown in Fig.~\ref{fig:3pidiagrams} and can be captured by combining $\eta_{2\pi}$ for the invariant mass of the $\pi^+\pi^-$ subsystem with Khuri--Treiman techniques~\cite{Khuri:1960zz} for the $\gamma^*\to 3\pi$ amplitude~\cite{Niecknig:2012sj,Schneider:2012ez,Hoferichter:2012pm,Danilkin:2014cra,Hoferichter:2014vra,Dax:2018rvs,Niehus:2021iin,Stamen:2022eda}, see Ref.~\cite{Hoferichter:2023bjm} for more details.  
 The results are summarized in the second panel of Table~\ref{tab:channels}, also exclusively in the $\Order(e^2)$ category.      

\emph{$\rho$--$\omega$ mixing}.---The interference of $\rho$ and $\omega$ arises in both the $2\pi$ and $3\pi$ channels, for the former there is an IB admixture of the $\omega$ to the dominant $\rho$ peak and vice versa. In the $2\pi$ case, this effect can be captured in a model-independent way by identifying the $\rho$--$\omega$ mixing parameter $\epsrw$ with the residue at the $\omega$ pole. Together with constraints on the threshold behavior, the resulting fit function is very rigid, with little freedom in the  line shape~\cite{Colangelo:2022prz}. In contrast, a signal for $\rho$--$\omega$ mixing in $e^+e^-\to 3\pi$ was first reported in Ref.~\cite{BABAR:2021cde}, yet based on a parameterization that violates the analyticity and unitarity properties of the amplitude. Here, we construct a 
 $\rho$--$\omega$ mixing contribution with sound analytic properties based on the coupled-channel framework of Ref.~\cite{Holz:2022hwz}, which predicts that the $\omega$ contribution in the dispersive formalism of Ref.~\cite{Hoferichter:2019gzf} be multiplied with 
 \beq
\label{3pi_rw}
g_\pi(s)=1-\frac{\gwg^2\epsrw}{e^2}\Pi_\pi(s), 
\eeq
where $\gwg$ parameterizes the $\omega$--$\gamma$ coupling and $\Pi_\pi$ denotes the vacuum-polarization function for a $\pi^+\pi^-$ pair.  In the narrow-width limit, the latter collapses to a $\rho$ propagator, with a coefficient $\gwg^2/\grg^2=10.9(1.2)$~\cite{Hoferichter:2017ftn,Holz:2022hwz} that comes close to the expected vector-meson-dominance enhancement factor $9$ compared to $2\pi$. The fact that the resulting sizable $\rho$--$\omega$ mixing effect in the $3\pi$ channel 
 cancels a substantial part of the $2\pi$ contribution can also be made plausible from narrow-resonance arguments, since the product of the two Breit--Wigner propagators, $P_V(s)=1/(s-M_V^2+i M_V\Gamma_V)$, $V=\rho,\omega$, as occurs naturally in the HVP function, decomposes as 
 \begin{align}
 P_\rho(s) P_\omega(s)&=\frac{1}{\mr^2-\mw^2-i\mr\Gr+i\mw\Gw}\notag\\
 &\times\Big(P_\rho(s)-P_\omega(s)\Big),
 \end{align}
 leading to terms in the $2\pi$ and $3\pi$ channel that tend to cancel, see Ref.~\cite{Hoferichter:2023bjm} for more details.  
 Finally, to separate $\epsrw$ onto $\Order(e^2)$ and $\Order(\delta)$ components, we rely on arguments from ChPT~\cite{Urech:1995ry,Bijnens:1996nq,Bijnens:1997ni}, which suggest that, after the $\gamma$--$\omega$ mixing diagram is removed in the bare cross section, $\epsrw$ should be primarily of $\Order(\delta)$ origin. The results for this class of contributions are given in the third panel of Table~\ref{tab:channels}.

\emph{Threshold effects}.---The final class of IB corrections arises from the mass differences of pion and kaon. For the pion, its mass is defined by the mass of the neutral pion in lattice-QCD calculations, which produces a large correction due to the low $2\pi$ threshold. Given this threshold enhancement, it is not sufficient to simply change the phase space, instead, we estimate the change incurred between neutral and charged pion mass using a dispersive representation of the pion vector form factor, calculating the pion-mass dependence of the phase shift in unitarized ChPT~\cite{Colangelo:2021moe,Niehus:2020gmf,Guo:2008nc,Omnes:1958hv}. Comparing one- and two-loop results, we observe a reasonable convergence pattern~\cite{Hoferichter:2022iqe}, in such a way that the dominant uncertainty arises from a low-energy constant at two-loop order. We cross-checked this result using a neutral-pion-mass phase shift obtained from solving Roy equations~\cite{Roy:1971tc,Ananthanarayan:2000ht,Garcia-Martin:2011iqs,Caprini:2011ky,Radinprep}, with input for the $\pi\pi$ scattering lengths at the neutral pion mass from two-loop ChPT~\cite{Niehus:2020gmf,Bijnens:1995yn,Colangelo:2001df}. The uncertainties given in Table~\ref{tab:channels} quantify the level of agreement between the two methods.

For the $\bar K K$ channel, a similar threshold enhancement occurs because the $\phi$ resonance is located very close to the $\bar K K$ threshold. In this case, we need to make sure that the decomposition of the kaon mass into $\Order(e^2)$ and $\Order(\delta)$ pieces matches the conventions in lattice QCD. Fortunately, the decomposition
\begin{align}
\label{mass_decomposition}
M_{K^\pm} &= \big(494.58-3.05_\delta + 2.14_{e^2}\big)\MeV, \notag\\ 
M_{K^0} &= \big(494.58+3.03_\delta\big)\MeV,
\end{align}
derived based on self energies that correspond to the elastic part of the Cottingham formula~\cite{Stamen:2022uqh,Cottingham:1963zz}, agrees well with typical schemes employed in lattice QCD (for which, in addition, the scale setting needs to be specified), in particular Ref.~\cite{Borsanyi:2020mff}. The results for all mass corrections are summarized in the fourth panel of Table~\ref{tab:channels}. We also considered pion-mass effects in the $3\pi$ channel, but in this case the corrections are strongly suppressed by phase space. Moreover, one could think that the pion-mass dependence of the $\omega$ width~\cite{Dax:2018rvs} might play a role, but this effect cancels out at the level of the integral.  

\nocite{Harlander:2002ur}

\begin{table}[t]
	\renewcommand{\arraystretch}{1.3}
	\centering
\begin{tabular}{clrrr}
\toprule
&& this work & Ref.~\cite{Borsanyi:2020mff} & Ref.~\cite{RBC:2018dos}\\\colrule
\multirow{2}{*}{SD} & $\Order(e^2)$ & $0.33(8)(8)(49)[51]$ & -- & -- \\
& $\Order(\delta)$ & $-0.04(4)(8)(49)[50]$ & -- & -- \\\colrule
\multirow{2}{*}{int} & $\Order(e^2)$ & $2.34(57)(47)(55)[92]$ & $-0.09(6)$ & $0.0(2)$ \\
& $\Order(\delta)$ & $0.02(33)(47)(55)[79]$ & $0.52(4)$ &  $0.1(3)$ \\\colrule
\multirow{2}{*}{LD} & $\Order(e^2)$ & $-1.97(63)(36)(12)[74]$ & -- & -- \\
& $\Order(\delta)$ & $1.48(44)(36)(12)[58]$ & -- & -- \\\colrule
\multirow{2}{*}{full} & $\Order(e^2)$ & $0.71(0.95)(0.90)(1.16)[1.75]$ & $-1.5(6)$  & $-1.0(6.6)$ \\
& $\Order(\delta)$ & $1.47(0.80)(0.90)(1.16)[1.67]$ & $1.9(1.2)$ & $10.6(8.0)$ \\
\botrule
\end{tabular}
\caption{Final results for the IB contributions in comparison to the lattice-QCD calculations from Refs.~\cite{Borsanyi:2020mff,RBC:2018dos}. The first error refers to the one propagated from Table~\ref{tab:channels}, the second one to the ambiguity in the $\phi$ residue, and the third to a generic $1\%$ correction to other channels not explicitly included in the analysis. The full $\Order(\delta)$ contribution can also be compared to the ChPT result $3.32(89)$ from Ref.~\cite{James:2021sor}, and the intermediate-window $\Order(e^2)$+$\Order(\delta)$ correction to $0.70(47)$ from Ref.~\cite{Ce:2022kxy} and $0.7(4)$ from Ref.~\cite{Giusti:2021dvd}. Previous lattice results for the full HVP integral include $1.1(1.0)_{e^2}$, $6.0(2.3)_{\delta}$ from Ref.~\cite{Giusti:2019xct} and $9.5(4.5)_{\delta}$ from Ref.~\cite{FermilabLattice:2017wgj}, see Ref.~\cite{Aoyama:2020ynm} for a review. All entries in units of $10^{-10}$.}  
	\label{tab:summary}
\end{table}

\emph{Uncertainty estimates}.---The errors given in Table~\ref{tab:channels} quantify the uncertainties in the determination of each particular effect, while the uncertainties of the sum are propagated in quadrature. The largest uncertainties arise in the $\rho$--$\omega$ mixing contribution in the $3\pi$ channel, largely due to the sensitivity to the assumed line shape of the amplitude~\cite{Hoferichter:2023bjm}, and from the pion mass difference in the $2\pi$ channel. However, before comparing to results from lattice QCD, we need to add uncertainties arising from subdominant channels or energy regions that we have not considered so far. We will estimate these uncertainties  as follows: first, we considered QED corrections to the $R$-ratio, but those scale as the expected $\Order\big(\frac{\alpha}{\pi}\big)=\Order(10^{-3})$~\cite{Harlander:2002ur}, and thus lead to negligible corrections 
$\lesssim 0.1\times 10^{-10}$ for the part of the HVP integral typically estimated with perturbative QCD. Second, an ambiguity arises in the identification of resonance couplings in the isospin limit. The first such case occurs in the $\bar K K$ channel, in which the residues of the $\phi$, $c_\phi$, in the fit to data reveal an IB effect of about $2\%$, $c_\phi=0.977(6)$ ($K^+K^-$) and $c_\phi=1.001(6)$ ($\bar K^0 K^0$). This IB effect is not included in Table~\ref{tab:channels}, since it is unclear how the isospin limit should be defined. Instead, we used the charged- (neutral-) kaon residue in the $K^+K^-$ ($\bar K^0 K^0$) channel. The opposite assignment produces a shift of 
\beq
\Delta a_\mu^{c_\phi}=0.90\times 10^{-10}
\eeq
in the $K^+K^-$ integral, and we will assign this value (and its decomposition onto the windows) as an additional source of uncertainty that should give an indication of missing effects in the exclusive channels that otherwise have been explicitly included.  
Finally, we need to estimate the impact of those channels not explicitly included, the largest of which being $e^+e^-\to 4\pi$. Absent any distinct resonance structures or threshold enhancements, we estimate the size of IB therein by a generic $1\%$ correction (motivated by the FSR effect in $2\pi$), which amounts to 
\beq
\Delta a_\mu^\text{missing channels} = 1.16\times 10^{-10},
\eeq
and this error could be scaled accordingly if a larger or smaller size of IB effects in the remainder were to be admitted. This final uncertainty is distributed onto the windows according to the isospin-conserving part of the missing channels. 

\begin{figure*}[t]
 \includegraphics[height=5.4cm]{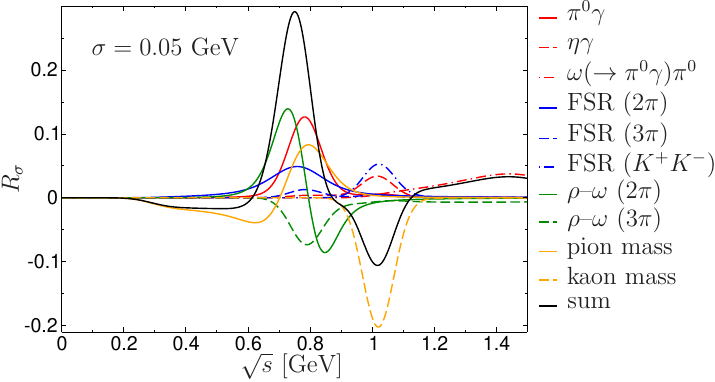}
 \includegraphics[height=5.4cm]{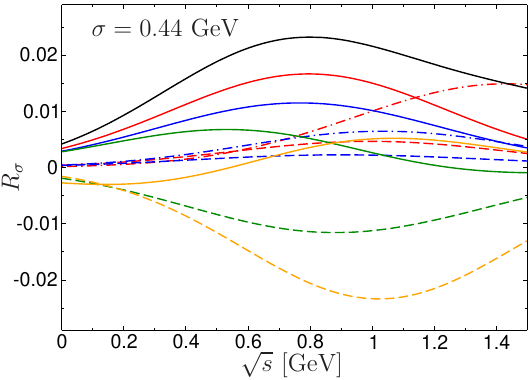}
 \caption{IB contributions to the $R$-ratio, smeared with Gaussian weights~\cite{ExtendedTwistedMassCollaborationETMC:2022sta}.}
 \label{fig:Gaussian}
\end{figure*}

\emph{Bottom line}.---Our final results are summarized in Table~\ref{tab:summary}, in comparison to the lattice-QCD computations from Refs.~\cite{Borsanyi:2020mff,RBC:2018dos}. In the SD window, the dominant uncertainty originates, as expected, from missing channels, since this part of the HVP integrals probes mostly the high-energy tail. Within uncertainties, both the $\Order(e^2)$ and $\Order(\delta)$ contributions are consistent with zero. The signal in the electromagnetic part mainly comes from the $\omega(\to\pi^0\gamma)\pi^0$ channel, due to its support around the $\rho(1450)$.  In the LD window, the higher channels only play a minor role, and we find a small positive (negative)  correction for strong IB (QED), both of which are the result of a cancellation among several sizable contributions.   
In the intermediate window, we see a larger $\Order(e^2)$ contribution than in lattice QCD, which becomes partly compensated for the full HVP integral. In the latter case, both $\Order(e^2)$ and $\Order(\delta)$ contributions agree within uncertainties, maybe with a slight excess in the QED correction. In view of the potential ambiguity at higher orders in ChPT in assigning $\epsrw$ to strong IB or QED, it is also instructive to consider the sum of both: in this case, IB in the full HVP contribution is compatible within uncertainties, while a larger effect in the intermediate window, entirely driven by the $\Order(e^2)$ contribution, survives.

The pattern with which the IB effects tend to cancel is illustrated in Fig.~\ref{fig:Gaussian}, using the smeared $R$-ratio~\cite{ExtendedTwistedMassCollaborationETMC:2022sta}
\beq
R_\sigma(\sqrt{s})=\int_0^\infty \diff\omega\frac{e^{-(\omega-\sqrt{s})^2/(2\sigma^2)}}{\sqrt{2\pi\sigma^2}}R_\text{had}(\omega). 
\eeq
For a fine resolution, $\sigma=0.05\GeV$ (left), the different channels produce sizable corrections: a negative correction around the $2\pi$ threshold, a net positive effect around the $\rho$--$\omega$ region, and again a negative correction in the vicinity of the $\phi$. Together with the $1/s^2$ weighting in Eq.~\eqref{amu_HVP}, the various contributions thus balance to a small overall IB correction. For a broader smearing as considered in Ref.~\cite{ExtendedTwistedMassCollaborationETMC:2022sta}, $\sigma=0.44\GeV$ (right), the distinct features are of course washed out, but also here a small net positive correction is obtained.

\emph{Conclusions}.---In this Letter, we systematically compiled the IB contributions to HVP that display some form of enhancement, via infrared singularities, threshold effects, or hadronic resonances, including, for the first time, both FSR and $\rho$--$\omega$ mixing in $e^+e^-\to 3\pi$. The resulting estimate of IB corrections is primarily motivated by the comparison to lattice QCD, but also of interest for a data-driven determination of quark-(dis)connected contributions~\cite{Boito:2022rkw,Boito:2022dry,Benton:2023dci}. Our central result is that the net IB corrections are small, both for QED and strong IB effects, as a result of cancellations among individually sizable contributions. The most noteworthy difference to lattice QCD arises for the QED contribution in the intermediate window, for which we obtain a non-negligible positive contribution, in contrast to essentially a null effect in Refs.~\cite{Borsanyi:2020mff,RBC:2018dos}.
Despite the modest size, it would be interesting to better understand the origin of the differences in the QED correction, but we emphasize that, even if they were to disappear in future lattice-QCD calculations, this would only increase the tension between lattice QCD and $e^+e^-\to\text{hadrons}$ data. From the size of the IB corrections as well as the sign of the difference to the estimates provided here, we thus conclude that the IB contributions in the lattice-QCD calculations can be ruled out as an explanation. 
 
We thank Laurent Lellouch and Antonin Portelli for valuable discussions.
Financial support by the SNSF (Project Nos.\ 200020\_175791, PCEFP2\_181117, and PCEFP2\_194272), the DFG through the funds provided to the Sino--German Collaborative
Research Center TRR110 ``Symmetries and the Emergence of Structure in QCD''
(DFG Project-ID 196253076 -- TRR 110),  and the Ram\'on y Cajal program (RYC2019-027605-I) of the Spanish MINECO is gratefully acknowledged.

\bibliography{amu}

\end{document}